\documentclass[aps,preprint,prl]{revtex4}

\usepackage{graphicx}
\usepackage{color}
\usepackage{amssymb}
\usepackage{amsmath} 
\usepackage{amstext} 
\usepackage{graphics}

\font\bb=msbm12

\newcommand{\bra}[1]{\left\langle#1\right|}
\newcommand{\ket}[1]{\left|#1\right\rangle}

\newcommand{\abs}[1]{\left|#1\right|}

\begin{document}

\preprint{ Version \today }

\title{ Charge-localization and isospin-blockade in vertical double quantum dots}

\author{David Jacob}
\affiliation{Departamento de F\'isica Aplicada
, Universidad de Alicante, San Vicente del Raspeig, 03690 Alicante, Spain}
\email{david.jacob@ua.es}
\author{Bernhard Wunsch}
\author{Daniela Pfannkuche}
\affiliation{I. Institute of Theoretical Physics, University of Hamburg, Jungiusstr. 9, D--20355 Hamburg, Germany}

\date{\today}

\begin{abstract}
\vspace{5mm} 
Charge localization seems unlikely to occur in two vertically coupled symmetric
quantum dots even if a small bias voltage breaks the exact isospin-symmetry of
the system. However for a
three-electron double quantum dot we find a strong 
localization of charges at certain vertically applied magnetic
fields. The charge localization is directly connected to new ground state
transitions between eigenstates differing only in parity. The transitions
are driven by magnetic field dependent Coulomb correlations between the
electrons and give rise to strong isospin blockade signatures in transport
through the double dot system.
\end{abstract}

\date{\today}


\maketitle

Quantum dot structures are excellent systems to investigate few and
many particle physics~\cite{Reimann:Oct2002,Kouwenhoven01:701,Rontani:2001}, due to the high
experimental control over the system parameters. In this context
double quantum dots are particularly interesting in two different
manners: as an implementation of quantum bits
(qubit)~\cite{Hu:May2000,Burkard:2000} and as a model system for
molecular binding under controlled
conditions~\cite{Rontani:2001,Partoens:Okt2001,Amaha183:01,Pi:Aug2001}.

In this letter we describe a new correlation effect in a vertically
coupled double quantum dot (DQD) in a perpendicular magnetic field,
which strongly changes the molecular binding and at the same time,
defines a two level system, that can be manipulated in a controlled
way and could serve as a qubit. This effect is manifest in the energy
spectrum and the transport properties of the DQD.  Sweeping the
magnetic field we find ground state crossings in a perfectly symmetric
DQD containing three electrons which~-- in contrast to the well known
crossings between states that differ in angular momentum and/or
spin~\cite{Imamura:Feb1999}~-- occur between states with same spin and
angular momentum. In contrast to a crossing between states that differ
in angular momentum and/or spin, that affects the lateral motion and
occurs in single dots already, the crossing discussed here involves a
transition in the parity of ground state, that characterizes the
vertical degree of freedom.  Therefore by slightly breaking the
symmetry between the two dots, e.g. by applying an infinitesimally
small voltage, the crossing turns into an anticrossing, resulting in
charge localization. Due to the charge localization, transport through
the DQD is strongly suppressed at the anticrossing. Reducing the
vertical degree of motion to an additional spin like degree of
freedom, the isospin, the strong suppression of transport can be
explained by an isospin blockade at the anticrossing in analogy to the
well known spin blockade~\cite{Weinmann95:984}.

We describe the DQD within the layer model~\cite{Imamura:Feb1999,Palacios:Jan1995}, that is applicable, if the
external potentials separate in a strong vertical and a considerably
weaker lateral component. We assume the in-plane confinement for the
electrons to be parabolic and circular symmetric.  Additionally a
magnetic field $B$ can be applied in the vertical direction.  The
in-plane motion of the electrons is then described in the effective
mass approximation by the Fock-Darwin-Hamiltonian $\hat{H}_{FD}$~\cite{Fock:Jan1928} and the Zeeman term $\hat{H}_{Z}$.
\begin{equation}
\label{FockDarwin}
  \hat{H}_{FD} + \hat{H}_Z= \frac{1}{2m^\ast} \left( \vec{p} + e \vec{A} \right)^2 + \frac{m^\ast\omega_0^2}{2} r^2 +
g^\ast \frac{\mu_B}{\hbar} B \hat{S}_z
\end{equation}
where $\omega_0$ is the strength of the parabolic confinement
potential, $m^\ast$ is the effective mass, $\mu_B$ the Bohr Magneton
and $g^\ast$ the effective Land\'e factor~\cite{material}.
The eigenstates of the in-plane motion are the Fock-Darwin states
$\ket{n,m}$ with the principal quantum number $n~\in~\mbox{\bb N}$ and
the angular momentum quantum number ($z$-component) $m~\in~\mbox{\bb
  Z}$.  The Hamiltonian (\ref{FockDarwin}) conserves the angular
momentum $\hat{L}_z$ as well as  the z-component of the spin $\hat{S}_z$ and the square of the spin $\hat{S}^2$,
described by $m$, $s_z$ and $s$ respectively.

The vertical motion is reduced to tunneling between two
$\delta$-sheets, labeled by the quantum number $\alpha~\in~\{+,-\}$.
$\alpha=\pm$ corresponds to the upper dot ($+$) or lower dot ($-$)
respectively.  In analogy with the real electron spin one can define a
spin operator algebra, where the $z$-component of the isospin,
$\hat{I}_z$ is given by $\alpha$~\cite{Palacios:Jan1995}.

The interdot
tunneling $\hat{H}_T~\ket{\pm}~=~t~\ket{\mp}$ which transfers
electrons between the two dots can be expressed by isospin operators:
\begin{equation}
  \hat{H}_T = t \left( \hat{I}_+ + \hat{I}_- \right) = 2 \, t \, \hat{I}_x
\end{equation}
with the real hopping parameter $t<0$~\cite{Imamura:Feb1999}.
$\hat{I}_{\pm}$ are the raising and lowering operators for the
$z$-component of the isospin, and $\hat{I}_x$ is its $x$-component.
The eigenstates of $\hat{I}_x$ and thus of $\hat{H}_T$ are the
symmetric and antisymmetric linear combinations of the isospin
eigenstates $\ket{\pm}$. Due to tunneling the electrons are
delocalized and the eigenstates of the Hamiltonian $\hat{H}_{FD} +
\hat{H}_T + \hat{H}_Z$ are no longer eigenstates of $I_z$. However in
case of symmetric dots the two layers are identical, so that the
isospin-parity $\hat{P}$ is conserved.  In case of more than one
electron inside the DQD, Coulomb interaction $\hat{\mathcal{V}}_c$
between the electrons has to be included such that the few-electron
Hamiltonian reads:
\begin{eqnarray}
  \label{hamiltonian}
  \hat{\mathcal{H}} 
  &=& \hat{\mathcal{H}}_{FD} + \hat{\mathcal{H}}_T + \hat{\mathcal{H}}_Z + \hat{\mathcal{V}}_c \\
  &=& \sum_{i=1}^{N_e} \left( \hat{H}_{FD}^{(i)} + \hat{H}_T^{(i)} + \hat{H}_Z^{(i)}\right)
  +\frac{e^2}{4\pi\epsilon\epsilon_0} \sum_{i<j} \hat{V}_c^{(i,j)} \notag.
\end{eqnarray}
Since Coulomb interaction is invariant under spatial and spin
rotations, total angular momentum  $\hat{\mathcal L}_z$ and total spin 
$\hat{\mathcal S}^2, \hat{\mathcal S}_z$ are still conserved
and are described by the quantum numbers $M$ and $S, S_z$ respectively. For
a symmetric DQD also the parity of the few electron eigenstates
$\hat{P} = 2^{N_e}\cdot \hat{I}_x^{(1)} \otimes \ldots \otimes \hat{I}_x^{(N_e)}$ 
is conserved, described by the quantum number $P~\in~\{+1,-1\}$.  Due to
Coulomb interaction the electrons are correlated. In a vertical double
quantum dot the Coulomb interaction can be divided into two parts:
$\hat{V}_c^{(i,j)} = \hat{V}_{intra}^{(i,j)} +
\hat{V}_{inter}^{(i,j)}$. The \emph{intradot} Coulomb interaction
$\hat{V}_{intra}^{(i,j)} = 1/r_{ij}$ describes the interaction between
electrons localized in the same dot, whereas the \emph{interdot}
Coulomb interaction $\hat{V}_{inter}^{(i,j)} = 1/( r_{ij}^2 + d^2
)^{1/2}$ describes the interaction between electrons localized in
different dots.  Here $r_{ij} = \abs{\vec{r}_i - \vec{r}_j}$ is the
lateral separation of two electrons $i$ and $j$ and $d$ is the
vertical separation between the dots. The Coulomb operator commutes
with the z-component of the total isospin $\hat{\mathcal{I}}_z$ but does not
commute with $\hat{\mathcal{I}}_x$ and accordingly $\hat{\mathcal{H}}_T$. The
commutator between Coulomb interaction and tunneling depends on the
difference between \emph{intradot} and \emph{interdot} Coulomb
interaction~\cite{Palacios:Jan1995} and vanishes only in the limit
$d\to~0$.

Increasing the vertical magnetic field effectively leads to a stronger
lateral confinement of the electrons and hence to an increase of the
Coulomb energy. Additionally intradot interaction increases faster
with increasing magnetic field than the interdot-interaction, that is
limited to $1/d$~\cite{Imamura:Feb1999}. This different scaling causes
magnetic field dependent correlations in the eigenstates. We show that
this can lead to a ground state crossing to fixed $M, S, S_z$ for
symmetric DQD and charge polarization in slightly asymmetric dots.

To take correlations into account we compute the eigenstates and the
corresponding eigenenergies by numerically diagonalizing the many-body
Hamiltonian (\ref{hamiltonian}), i.e. we expand the eigenstates in a
finite basis of Slater determinants~\cite{basis}.

\begin{figure}
  \includegraphics{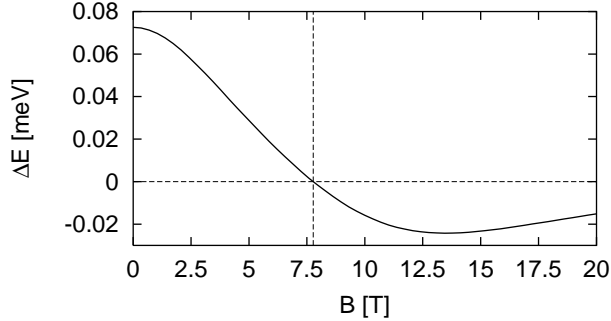} 
  \caption{
    Energy difference of the lowest two eigenstates to $\hat{\mathcal{H}}$ 
    ($M=-5$ and $S=S_z=3/2$) as a function of the magnetic field B.
    $t = -0.059~{\rm meV}$, $\hbar \, \omega_0 = 2.96~{\rm meV}$ 
    and $d = 19.6~{\rm nm}$. The crossing takes place at B = 7.77~T
    (dashed vert. line).}
  \label{fig:energy-diff}
\end{figure}

\begin{figure}
 \includegraphics[scale=1, angle=0]{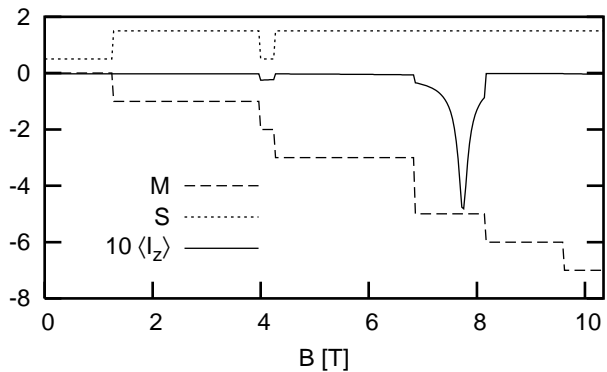}
  \caption{
    Angular momentum $M$, total spin $S$ and expectation value of the z-component of the isospin $\langle \mathcal{I}_z \rangle$
    for the three electron ground state to $\hat{\mathcal{H}} + \hat{\mathcal{V}}_z$.
    $\hat{\mathcal{V}}_z= 5.9*10^{-4} \, \hat{\mathcal{I}}_z~{\rm meV}$ 
    represents a slight asymmetry between the dots. The peak in $\langle \hat{\mathcal{I}}_z \rangle$ illustrates the charge localization that corresponds to the parity crossing (see text). Other parameters are the same as in Fig.~\ref{fig:energy-diff}.
    }
  \label{fig:localization}
\end{figure}

\begin{figure}
  \includegraphics{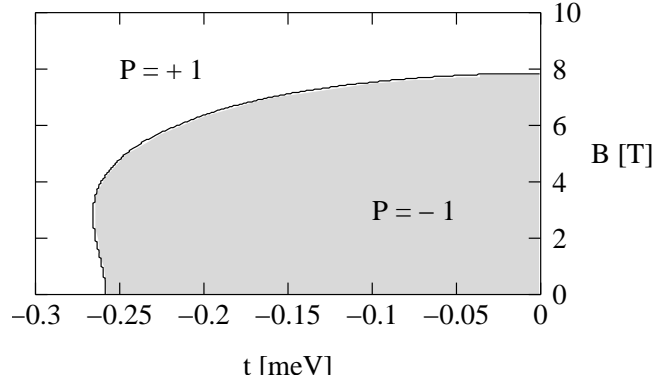}
  \caption{
    Dependence of the parity $P$ for angular momentum
    $M=-5$ and $S=3/2$ on magnetic field $B$ and tunneling $t$.
    The solid line indicates where the crossing
    between the parity eigenstates takes place. 
    Other parameters are the same as in Fig.~\ref{fig:energy-diff}.
    }
  \label{fig:parity-diagram}
\end{figure}

Calculating the magnetic-field dependence of the energy spectrum for
three electrons inside a symmetric DQD to angular momentum $M=-5$
and spin $S=S_z=3/2$, we find a crossing between the two energetically
lowest states as illustrated in Fig.~\ref{fig:energy-diff}.
Since the crossing states only differ in parity, the accidental 
crossing converts into an anticrossing if the parity conservation 
is broken by a slight asymmetry between the dots leading to two
strongly charge-polarized states.
For specific parameters the parity crossing and hence the 
charge-polarization found for this subspace of quantum numbers 
becomes visible in the ground state (GS) as illustrated 
in Fig.~\ref{fig:localization}.
For other parameters the parity crossing
will either arise in excited states or even dissappear completely as
shown in Fig.~\ref{fig:parity-diagram}. The asymmetry between the dots
can be either intrinsic or caused by a small bias voltage, as it is
applied in transport experiments~\cite{Ancilotto:2003}. We
model the asymmetry between the dots by adding the term
$\hat{\mathcal{V}}_z = V_z \cdot \hat{\mathcal{I}}_z$ to the Hamiltonian
(\ref{hamiltonian}) where $V_z$ is the energy difference between upper
and lower dot for a single electron.  While the ground state is nearly
unpolarized for general magnetic field strengths,
Fig.~\ref{fig:localization} shows a strongly polarized ground state at
the magnetic field where the anticrossing occurs. The minimal value of
$\left\langle \mathcal{I}_z \right \rangle =-0.5$ corresponds to two electron
charges in the lower dot and one in the upper.  
Thus we find the
astonishing effect that electrons become localized in one of the dots
by simply changing the vertical magnetic field.
It is important to note that the strength of the asymmetry (i.e. $V_z$)
only determines the width of the localization dip in Fig.~\ref{fig:localization}
but even for arbitrarily small asymmetries the ground state is strongly polarized at
the anticrossing with $\langle \mathcal{I}_z \rangle =-0.5$.
Since
$[\hat{\mathcal{L}}_z,\hat{\mathcal{V}}_z] =[\hat{\mathcal{S}}^2,\hat{\mathcal{V}}_z] =
[\hat{\mathcal{S}}_z,\hat{\mathcal{V}}_z] = 0$, $\hat{\mathcal{V}}_z$ couples
only states with same total angular momentum and total spin. Therefore
a similar effect does not occur in the well known ground state
crossing between states that differ in $M$ and/or $S$~\cite{Imamura:Feb1999}.

In the following we study the reported parity crossing in the GS of a
symmetric DQD in more detail.  Without tunneling $I_z$ is conserved
and since both dots are identical the ground state will be two-fold
degenerate with $I_z=\pm0.5$.  Switching on tunneling their degeneracy
is lifted and the GS splits in two non-degenerate parity
eigenstates $\ket{P=\pm1}$ because of their different occupations of
symmetric and antisymmetric orbitals.  In particular Fig.~\ref{fig:energy-diff} illustrates that for magnetic fields $B < 7.8$~T
$\ket{P=-1}$ is favored by tunneling i.e. it
has a higher occupation of symmetric orbitals than $\ket{P=+1}$.
However due to magnetic-field dependent correlations the occupation of symmetric orbitals decreases for 
$\ket{P=-1}$ but increases for $\ket{P=+1}$, so that by increasing the magnetic field
finally $\ket{P=+1}$ becomes the GS.
Fig.~\ref{fig:parity-diagram} shows
the parity as a function of tunneling and 
external magnetic field
for the subspace $M=-5$ and
spin $S=S_z=3/2$.
The crossing exists from zero tunneling up to 
$t\approx0.27~{\rm meV}$, which suggests to
treat the tunneling $t$ as a small perturbation.
For small tunneling (tunneling much smaller than 
the energy spacing between degenerate ground state
and first excited state at $t=0$) the parity eigenstates 
are to first order perturbation theory given by
$\ket{P=\pm1}\approx(\ket{I_z=\frac{1}{2}} \pm\ket{I_z=-\frac{1}{2}})/\sqrt{2}$. 
and their energy splitting is 
$2\bra{I_z=\frac{1}{2}}\hat{\mathcal H}_T\ket{I_z=-\frac{1}{2}}(B)$.  
As indicated this matrix element depends on the magnetic field 
due to the magnetic-field dependent correlations present in 
the states $\ket{I_z=\pm\frac{1}{2}}$.  
To first order the crossing occurs at $B=7.85$~T 
where the matrix element vanishes, and is independent 
of $t$ in good agreement with the exact results for
small tunneling (see Fig.~\ref{fig:parity-diagram}).
For strong tunneling however higher order effects 
(coupling to higher states) come into play causing 
the crossing to disappear for $t >0.27~{\rm meV}$.
Breaking the vertical symmetry of the DQD the two parity eigenstates
are coupled and the parity-crossing converts into an anticrossing,
thereby lifting their accidental degeneracy by an amount $V_z$.
At the anticrossing
the eigenstates are approximately given by $I_z=\pm\frac{1}{2}$ and
are thus strongly charge polarized.
We want to note that the parity-crossing and the related strong charge-polarization is 
not restricted to the total angular momentum and total spin chosen 
here but also occurs for other sets of quantum numbers.
Furthermore a symmetric DQD containing any odd number
of electrons has a degenerate ground state at $t=0$ and similar parity
crossings are expected for higher odd numbers of electrons in the DQD
(the ground state of an even number of electrons at $t=0$ has $I_z=0$
and is non-degenerate).

\begin{figure}
  \includegraphics{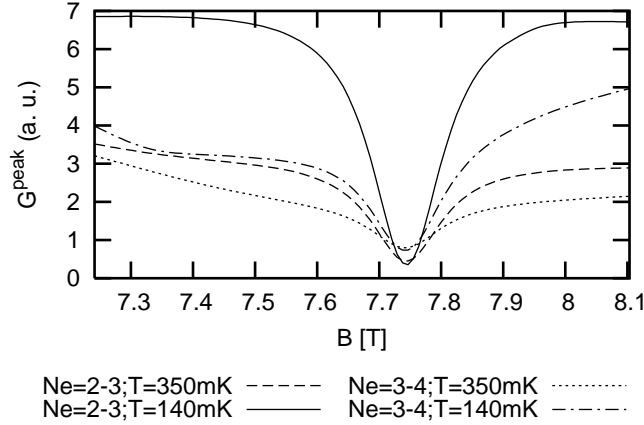}
  \caption{
    Calculated height of third and fourth conductance peak (transition
    from $N_e=2$ to $N_e=3$ or from $N_e=3$ to $N_e=4$)) as a function
    of the magnetic field for two temperatures.  Parameters:
    $V_{sd}=12\mu$V; T = 140~mK . Other parameters are the same as
    in Fig.~\ref{fig:energy-diff} and \ref{fig:localization}.  In
    particular the asymmetry between the lower and upper dot
    $V_z=10^{-2} |t| \approx 0.6 \mu\rm{eV}$.  }
  \label{fig:conductance}
\end{figure}

The polarization of the three-electron GS can be detected in
a transport experiment through the DQD~\cite{Austing206:1998,Pi:Aug2001,Amaha183:01,Ancilotto:2003}.  
If a small transport voltage, $V_{SD}$, accross the DQD is applied~\cite{transport}
at constant magnetic field, the conductance $G$ has a
peak structure as a function of the gate voltage. The height of the
conductance peaks $G^{peak}$ corresponding to the transitions between two and
three electrons or three and four electrons inside the DQD are shown
as a function of the magnetic field and for two different temperatures
in Fig.~\ref{fig:conductance}.  A comparison with Fig.~\ref{fig:localization} shows that the current through the DQD is
suppressed at the magnetic field, where the three-electron GS
becomes polarized. We want to point out that since the asymmetry
between the dots is weak only the two lowest three-electron states are
polarized (in opposite direction) whereas the other states and in
particular the two and four electron ground states are unpolarized (in contrast to Ref.~\cite{Pi:Aug2001}).
In our calculations we assume that transport is described by
sequential tunneling processes in and out of the many-particle
eigenstates of the isolated DQD.
This is a good approximation for weak tunnel contacts between the
reservoirs and the DQD, i.e. the tunneling strength to the external
reservoirs is smaller than the interdot tunneling and the finite
lifetime broadening of the DQD states is smaller than temperature~\cite{Beenakker91:1646}. 
For the tunneling events a transition rate can be calculated, which we
call $T^+$ ($T^-$) for a transition caused by a tunneling event
through the upper (lower) barrier. In the following we discuss the
transition between two and three electrons in the dot, but the
arguments are equally valid also for the next conductance peak.

Assuming that an electron in the upper (lower) reservoir can only
tunnel into the upper (lower) dot, the transition rate $T^+$ between a
two particle state and a three particle state is proportional to the
spectral weight $T^+_{N_e=3 \to N_e=2} \propto \sum_{n,m,\sigma}
\abs{\bra{N_e=2} d_{n m + \sigma} \ket{N_e=3} }^2$~\cite{Pfannkuche98:1255}, where $d_{n m +
  \sigma}$ denotes the annihilation operator for the orbital $\ket{n m
  + \sigma}$ in the upper dot~\cite{basis}, 
similarly $T^-_{N_e=3 \to
  N_e=2} \propto \sum_{n,m,\sigma} \abs{\bra{N_e=2} d_{n m - \sigma}
  \ket{N_e=3}}^2$.  Due to the small
transport voltage only the two transport channels that include the
unpolarized two-electron GS and one of the polarized
three-electron states (GS and first excited state) lie
within the transport window.  Higher channels only contribute due to
the finite temperature and can be further suppressed by lowering the
temperature.  For both transport channels that include polarized 
three-electron states one of the transition rates either for the tunneling
in or tunneling out process is isospin-blocked. $T^-$ ($T^+$) is
suppressed, if the three electron state has two electrons localized in
the upper (lower) dot. For a current to flow through the DQD both
tunneling processes are necessary, which is expressed by the effective
tunneling rate $\Gamma\propto \frac{T^-\,T^+}{T^-+T^+}$~\cite{Klimeck94:2316}.  
Therefore the current is strongly reduced due
to an isospin blockade of both channels.  Away from the crossing the
three-electron states are no longer polarized so that the transition
through both barriers is possible.

We conclude: 
Magnetic-field
dependent Coulomb correlations
affect the eigenstates' tunneling energies 
differently depending on their parity
leading to additional magnetic-field
induced level-crossings between states with 
different parity but same angular momentum 
and spin in a perfectly symmetric DQD.
The magnetic-field dependent charge-polarization 
in symmetry-broken DQDs is due to an anticrossing 
of two eigenstates with different parity.
The charge polarization takes also place in the ground state and is
detectable in a transport experiment through the DQD as it leads to an
isospin blockade at the magnetic field where the polarization occurs.
The resulting polarized eigenstates $\ket{\frac{1}{2}}$ and
$\ket{-\frac{1}{2}}$ can be seen as a qubit which can be
switched by the applied bias voltage. 
A controlled superposition of the two states can then be achieved by
adjusting the magnetic field.  
The parity crossing and the related
magnetic-field dependent charge polarization appear also for an odd
number of electrons greater than three and in different subsets of
quantum numbers.

We thank Michael Tews for his intensive help with the transport
calculations and acknowledge financial support by the Deutsche
Forschungsgemeinschaft via SFB 508.  

{\small \bibliographystyle{apsrev} \bibliography{double_dot} }

\begin{thebibliography}{20}
\bibitem[{mat(We assume GaAs as host material: $m^\ast = 0.067 \, m_e$ and
  effective Land\'e $g^\ast=-0.44$.)}]{material}
 \bibinfo{year}{We assume GaAs as host material: $m^\ast = 0.067 \, m_e$ and
  effective Land\'e $g^\ast=-0.44$.}
\bibitem[{bas(A suitable single-particle basis for our problem form the
  Fock-Darwin states $\ket{n,m}$ with isospin $\ket{\alpha}$ and real spin
  $\ket{\sigma}$. Since $M$, $S$ and $S_z$ are conserved, we can diagonalize in
  subspaces to constant $M$, $S$ and $S)}]{basis}
 \bibinfo{year}{A suitable single-particle basis for our problem form the
  Fock-Darwin states $\ket{n,m}$ with isospin $\ket{\alpha}$ and real spin
  $\ket{\sigma}$. Since $M$, $S$ and $S_z$ are conserved, we can diagonalize in
  subspaces to constant $M$, $S$ and $S_z$.}
\bibitem[{tra($e V_{SD} \ll \Delta \mu$ where $\Delta \mu$ describes the change
  of the chemical potential inside the DQD when an electron is added to the
  DQD.)}]{transport}
 \bibinfo{year}{$e V_{SD} \ll \Delta \mu$ where $\Delta \mu$ describes the
  change of the chemical potential inside the DQD when an electron is added to
  the DQD.}
\end{thebibliography}

\end{document}